\documentclass[a4paper]{JHEP3}
\usepackage{mathrsfs}
\usepackage{amssymb}

\usepackage{graphicx}

\title{Fermion Families from Two Layer Warped Extra Dimensions}

\author{Zhi-qiang Guo$^{1}$ and Bo-Qiang
Ma$^{1}$\thanks{Email:mabq@phy.pku.edu.cn}  \\
$^{1}$School of Physics and State Key Laboratory of Nuclear Physics
and Technology, Peking University, Beijing 100871, China }

\abstract{In extra dimensions, the quark and lepton mass hierarchy
can be reproduced from the same order bulk mass parameters, and
standard model fermion families can be generated from one generation
in the high dimensional space. We try to explain the origin of the
same order bulk mass parameters and address the family replication
puzzle simultaneously. We show that they correlate with each other.
We construct models that families are generated from extra
dimensional space, and in the meantime the bulk mass parameters of
same order emerge naturally. The interesting point is that the bulk
mass parameters, which are in same order, correspond to the
eigenvalues of a Schr\"{o}dinger-like equation. We also discuss the
problem existing in this approach.}


\begin{document}

\section{Introduction}\label{sec:1}

In recent years, extra space dimensions have played an important
role in particle physics, gravity and cosmology. Many old problems
get elegant solutions in extra space dimensional background with new
perspectives \cite{Arkani-Hamed:1998,Randall:1999}. The extra space
dimensions have been used to address the problems in flavor physics:
why the masses of quarks and leptons distribute in a large range and
have obvious hierarchy structure~(the fermion mass hierarchy
puzzle); why the heavier generations replicate the lightest
generation with the almost same properties~(the fermion family
replication puzzle). There have been very interesting progress that
both of them get answers in several different approaches
\cite{Libanov:2001,Dobrescu:2001,Nernov:2002,Kaplan:2001,Csaki:2005}.

In extra dimension, the fermion mass hierarchy has geometrical
origin. It arises from the small overlap of wave functions in the
extra space dimensions. The hierarchy mass structure can be
reproduced with the bulk mass parameters of the same order in
5-dimension warped space~(Randall-Sundrum model) successfully~(for
several numerical examples, see
\cite{Hosotani:2006-1}).\footnote{For definiteness, our following
discussions will be based on the concise numerical example given in
\cite{Hosotani:2006-2}, and we hope that our discussions can apply
to other examples by some modification.} However, a new puzzle
arises naturally: why the 5-dimension bulk mass parameters are of
the same order, or whether we can give an explanation to the origin
of the same order 5-dimension bulk mass parameters. It is easy to
understand that this new puzzle correlates closely with the fermion
family replication problem. One bulk mass parameter stands for one
fermion family or one fermion flavor, so to explain the origin of
the same order bulk mass parameter is doing the same thing to
explain the family replication problem.

This new puzzle can be addressed in the approach that the three
generations of standard model~(SM) can be generated from one
generation in the high dimensional space. This approach has been
adopted in several papers \cite{Libanov:2001,Nernov:2002}, in which
the 6-dimension spacetime is reduced to 4-dimension spacetime
directly. It is found that the SM families correspond to the zero
modes of the high dimensional equation of motion. In these
approaches, one missed the chance to give answers for the origin of
the same order bulk mass parameters in 5-dimensions. In the present
paper, we will adopt an alternative approach. We suggest a
6-dimension metric Ansatz of special two layer warped structure,
\begin{equation}
\label{introduce} ds^2 = B(z)^2\left[A(y)^2\left( \eta_{\mu \nu}
dx^{\mu} dx^{\nu} + d y^2\right)+ d z^2\right].
\end{equation}
As we will show below, with the help of the special structure of
this two layer warped metric, we can reduce the 6-dimension (4+1)+1
spacetime to 5-dimension 4+1 spacetime at the first step. We found
that the induced 5-dimension equation of motion for fermions is
similar to that analyzed in \cite{Hosotani:2006-2}, while the bulk
mass parameter in this induced 5-dimension equation corresponds to
the eigenvalue of a 1-dimension Schr\"{o}dinger-like equation, which
is conformed by the fermion wave function in the sixth space
dimension. Yet, because the induced 5-dimension equation of motion
is similar to that analyzed in \cite{Hosotani:2006-2}, we might
expect that this induced 5-dimension equation  can be applied in
this model by some modification. Hence we reduce the problem of the
same order 5-dimension bulk mass parameter to the eigenvalue problem
of a second order differential equation. Because the eigenvalues of
a 1-dimension Schr\"{o}dinger-like equation are of the same order
generally, we see that the same order 5-dimension mass parameters
emerge naturally. If we further reduce the 5-dimension spacetime to
4-dimension spacetime, with the help of the model
\cite{Hosotani:2006-2}, the same order 5-dimension bulk mass
parameters will produce the hierarchy structure of the 4-dimension
physical fermion mass. By this new approach, we can address the new
puzzle suggested above. We give an explanation for the origin of the
same order bulk mass parameters. Because one 5-dimension bulk mass
parameter corresponds to one family, we provide an answer for the
origin of the families simultaneously. However, there exist some
problems in this approach, we will discuss these problems in detail
and suggest methods to bypass them.

According to the approach suggested above, we construct a model that
the 5-dimension bulk mass parameters of the same order emerge
naturally, and hence the standard model families are generated
simultaneously. In fact, we can show that the 5-dimension bulk mass
parameters correspond to the eigenvalues of a second order ordinary
differential equation of the Sturm-Liouville type (similar to the
Schrodinger equation in 1-dimension). Therefore, like the energy
spectrum of Hydrogen atom, the bulk mass parameters or the families
can be generated from the eigenvalues of Schr\"{o}dinger-like
equation with proper boundary conditions. The details will be
introduced in Sec.~\ref{sec:2}. Several examples are given in
Sec.~\ref{sec:3}. We give further discussions and conclusions in
Sec.~\ref{sec:4} and Sec.~\ref{sec:5}.

\section{The Setup}\label{sec:2}

In this section, we introduce the setup in detail. We start with the
action of a bulk Dirac fermion in six dimension spacetime. The
metric Ansatz for spacetime takes the form
\begin{equation}
\label{ansatz1} ds^2 = B(z)^2\left[A(y)^2\left( \eta_{\mu \nu}
dx^{\mu} dx^{\nu} + d y^2\right)+ d z^2\right],
\end{equation}
where $\eta_{\mu \nu}=\mathrm{diag}(-1,1,1,1)$. Note that the metric
has the special two layer warped structure. By this special metric
structure, one can reduce the 6-dimension spacetime to 5-dimension
spacetime at the first step, then further reduce the 5-dimension
spacetime to 4-dimension physical spacetime, as we will introduce in
detail below. We suppose that the extra dimensions both are
intervals.\footnote{The conventional way in extra dimensions is to
suppose the extra dimensions as orbifolds, and that the boundary
conditions are determined to be the Israel's junction conditions
\cite{Israel:1966}, as adopted in \cite{Randall:1999}. However, as
suggested in \cite{Csaki:2005-1}, the interval approach is more
convenient in some cases, and it can produce more general boundary
conditions. In our work, we found that it is necessary to adopt the
interval approach, at least when we deal with the boundary
conditions in the sixth space dimension. Of course, it is also more
convenient. More about the interval approach, see
\cite{Csaki:2005-2}.}

The bulk action for this fermion is given by the usual form,
\begin{equation}
        \label{action1}
S = \int d^4 x dy dz \sqrt{-g} \left\{ \frac{i}{2} \left[\bar{\Psi}
\, e_a^M \Gamma^a D_M \Psi - D_M \bar{\Psi} \, e_a^M \Gamma^a \Psi
\right] -i \, m \bar{\Psi} \Psi\right\},
\end{equation}
where $e_a^M$ is the sechsbien, and $D_M=\partial_M + \frac{1}{2}
\omega^{ab}_M
\Gamma_{ab},\Gamma_{ab}=\frac{1}{4}[\Gamma_a,\Gamma_b]$ is the
covariant derivative of spinor in curved spacetime. $a$ and
$M=0,1,2,3,5,6$ stand for the flat spacetime indices and the curved
spacetime indices respectively. The Dirac equation in 6-dimension
spacetime requires $m$ to be a real number. We choose the gamma
matrices representation as follows,
\begin{eqnarray}
\label{Eq3} \Gamma^\mu &=& \left(
\begin{array}{cc}
0 & \gamma^\mu \\
\gamma^\mu & 0
\end{array}
\right), ~\Gamma^5 = \left(
\begin{array}{cc}
0 & \gamma^5 \\
\gamma^5 & 0
\end{array}
\right), ~\Gamma^6 = \left(
\begin{array}{cc}
\mathbf{1}_4 & 0 \\
0 & -\mathbf{1}_4
\end{array}
\right), \nonumber \\
\gamma^0 &=& \left(
\begin{array}{cc}
0 & -\mathbf{1}_2 \\
\mathbf{1}_2 & 0
\end{array}
\right), ~\gamma^i = \left(
\begin{array}{cc}
0 & \sigma^i \\
\sigma^i & 0
\end{array}
\right), ~\gamma^5 = \left(
\begin{array}{cc}
\mathbf{1}_2 & 0 \\
0 & -\mathbf{1}_2
\end{array}
\right), ~~i=1,2,3,
\end{eqnarray}
where $\mu=0,1,2,3$ and $\sigma^i$ are the usual Pauli matrices.

By use of the metric Ansatz, the action reduces to
\begin{eqnarray}
        \label{action2}
S &=& S_4+S_5+S_6 -\int d^4 x dy dz \sqrt{-g}i \, m \bar{\Psi} \Psi, \\
S_4 &=& \int d^4 x dy dz \sqrt{-g} \left\{ \frac{i}{2} B(z)^{-1}
A(y)^{-1} \left[\bar{\Psi} \,\Gamma^\mu
\partial_\mu \Psi -
\partial_\mu \bar{\Psi} \Gamma^\mu \Psi\right] \right\}, \nonumber \\
S_5 &=& \int d^4 x dy dz \sqrt{-g} \left\{ \frac{i}{2} B(z)^{-1}
A(y)^{-1} \left[\bar{\Psi} \,\Gamma^5
\partial_5 \Psi -
\partial_5 \bar{\Psi} \Gamma^5 \Psi\right]\right\}, \nonumber \\
S_6 &=& \int d^4 x dy dz \sqrt{-g} \left\{ \frac{i}{2} B(z)^{-1}
\left[\bar{\Psi} \,\Gamma^6 \partial_6 \Psi -
\partial_6 \bar{\Psi} \Gamma^6 \Psi\right]\right\}.\nonumber
\end{eqnarray}
Varying the action with respect to $\bar{\Psi}$, we obtain the
equation of motion,
\begin{eqnarray}
\label{solution4} A(y)^{-1} \left\{\Gamma^\mu
\partial_\mu \Psi + \Gamma^5
\left(\partial_5 + 2 A^{-1}A'\right) \Psi\right\} + \Gamma^6
\left(\partial_6 + \frac{5}{2} B^{-1}\dot{B}\right) \Psi - m \, B \,
\Psi=0,
\end{eqnarray}
where $A'= \frac{d A(y)}{d y}, \dot{B}= \frac{d B(z)}{d z},
\partial_5=\partial_y,\partial_6=\partial_z$, and the boundary term,
\begin{eqnarray}
\label{solution5} \delta S_{\mathrm{bound}}&=& \delta
S_{5\,\mathrm{bound}}+\delta
S_{6\,\mathrm{bound}}, \\
\delta S_{5\,\mathrm{bound}} &=& -\frac{i}{2} \int d^4x dz \left[
\sqrt{-g}~B^{-1}A^{-1}\delta \bar{\Psi} \Gamma^5 \, \Psi \right]_{L}^{L'},\nonumber \\
\delta S_{6\,\mathrm{bound}}&=&-\frac{i}{2} \int d^4x dy
\left[\sqrt{-g}~B^{-1}\delta \bar{\Psi} \Gamma^6 \, \Psi
\right]_{R}^{R'},\nonumber
\end{eqnarray}
where we denote by $[X]_0^L$ the quantity $X_{|L}-X_{|0}$. Denoting
$\Psi=\left(
\begin{array}{c}
\chi_{1}\\
\chi_{2}
\end{array}\right)$, where $\chi_{1}$ and $\chi_{2}$ are four-component Dirac spinors, we rewrite Eq.~(\ref{solution4}) as
\begin{eqnarray}
\label{solution6} A(y)^{-1} \left\{\gamma^\mu
\partial_\mu \chi_2 + \gamma^5
\left(\partial_5 + 2 A^{-1}A'\right) \chi_2\right\} +
\left(\partial_6 + \frac{5}{2}
B^{-1}\dot{B}\right) \chi_1 - m \, B \, \chi_1 &=& 0,\\
\label{solution7} A(y)^{-1} \left\{\gamma^\mu
\partial_\mu \chi_1 + \gamma^5
\left(\partial_5 + 2 A^{-1}A'\right) \chi_1\right\} +
\left[-\left(\partial_6 + \frac{5}{2} B^{-1}\dot{B}\right)\right]
\chi_2 - m \, B \, \chi_2 &=& 0.
\end{eqnarray}

Now we make the conventional Kluza-Klein~(KK) decomposition. We
expand $\chi_{1}$ and $\chi_2$ with spinor $\psi(x^\mu,y)$ in
5-dimension spacetime as
\begin{eqnarray}
\label{solution9} \chi_1(x^\mu,y,z)= \sum_n \widehat{F}_n(z)\,
\psi_{n} (x^\mu,y),~\chi_2(x^\mu,y,z)= \sum_n \widehat{G}_n(z)\,
\psi_{n} (x^\mu,y),
\end{eqnarray}
in which $\psi_{n} (x^\mu,y)$ conforms to the Dirac equation in
5-dimension spacetime,
\begin{eqnarray}
\label{solution9} A(y)^{-1}\left\{\gamma^\mu
\partial_\mu \psi_{n} (x^\mu,y) + \gamma^5
\left(\partial_5 + 2 A^{-1}A'\right) \psi_{n}
(x^\mu,y)\right\}-\lambda_n \psi_{n} (x^\mu,y) = 0.
\end{eqnarray}
As in 6-dimension spacetime, the Dirac equation in 5-dimension
spacetime requires $\lambda_n $ to be real numbers. This can be
verified by multiplying the two sides of Eq. (\ref{solution9}) by
$\bar{{\psi_{n}}}(x^\mu,y)$. Here we note that it is critical that
$\lambda_n $ must be real numbers, as it will be obvious in the
following discussions. With the help of Eq. (\ref{solution9}), Eqs.
(\ref{solution6})-(\ref{solution7}) can be solved by the following
Ansatz,
\begin{eqnarray}
\label{solution10}
\left(\frac{d}{d z}+\frac{5}{2} B^{-1}\dot{B}\right)\widehat{F}_n(z)-m B \widehat{F}_n(z) + \lambda_n \widehat{G}_n(z)=0,\\
\label{solution11} \left(\frac{d}{d z}+\frac{5}{2}
B^{-1}\dot{B}\right)\widehat{G}_n(z)+m B \widehat{G}_n(z) -
\lambda_n \widehat{F}_n(z)=0.
\end{eqnarray}
These equations can be simplified further by the transformations,
\begin{eqnarray}
\label{solution12}
\widehat{F}_n(z)=B(z)^{-\epsilon}F_n(z),~\widehat{G}_n(z)=B(z)^{-\epsilon}
G_n(z),
\end{eqnarray}
in which $\epsilon=\frac{5}{2}$. Then we obtain the equations
\begin{eqnarray}
\label{solution13}
\left(\frac{d}{d z}- m B\right)F_n(z) + \lambda_n G_n(z)=0,\\
\label{solution14} \left(\frac{d}{d z}+ m B \right)G_n(z) -
\lambda_n F_n(z)=0.
\end{eqnarray}
For a zero mode ($\lambda=0)$, these bulk equations decouple and are
easy to be solved. The solutions are given by
\begin{eqnarray}
\label{ZeroMode} F_{0}(z)=\frac{1}{\sqrt{l
N_0}}\mathrm{exp}\left(\int_{z_0}^{z}mB(\zeta)d\zeta\right)~\mathrm{or}~0,
G_{0}(z)=\frac{1}{\sqrt{l\widetilde{N}_0}}\mathrm{exp}\left(-\int_{z_0}^{z}mB(\zeta)d\zeta\right)~\mathrm{or}~0.
\end{eqnarray}
We introduce $l$ of the length dimension in order to make the
normalization constants to be dimensionless. It will become explicit
in examples in the next section. For the massive modes, we can
combine the first order differential equation to obtain second order
equations
\begin{eqnarray}
\label{solution15} \frac{d^2}{d z^2}F_n(z)
+\left[-m \dot{B}-m^2 B^2\right]F_n(z)+\lambda_n^2 F_n(z)=0,\\
\label{solution16} \frac{d^2}{d z^2}G_n(z)+\left[m \dot{B}-m^2
B^2\right]G_n(z)+\lambda_n^2 G_n(z)=0.
\end{eqnarray}
Rewriting them in another form, we see that they are similar to the
one dimensional Schr\"{o}dinger equations
\begin{eqnarray}
\label{solution17} -\frac{d^2}{d z^2}F_n(z)
+V(z)F_n(z)&=&\lambda_n^2 F_n(z),\\
\label{solution18} -\frac{d^2}{d z^2}G_n(z)+\widetilde{V}(z) G_n(z)
&=&\lambda_n^2 G_n(z),
\end{eqnarray}
with potentials
\begin{eqnarray}
\label{solution19} V(z)=m \dot{B}+m^2 B^2,~\widetilde{V}(z)=-m
\dot{B}+m^2 B^2.
\end{eqnarray}

In the following discussions we will illuminate that such a setup
gives answers to the puzzle we proposed in the introduction.
According to the setup above, we realize an interesting fact that
Eq.~(\ref{solution9}) is similar to the equation analyzed in the
model \cite{Hosotani:2006-2} if we choose $A(y)$ to be a slice of
the anti-de Sitter~(AdS) metric, i.e., the RS spacetime. The
differences are that there is gauge field background in the model
\cite{Hosotani:2006-2}, and that in that work the extra dimension is
adopted to be an orbifold. However, for fermions, the equations of
motion in these two cases are of almost similar features, and the
gauge field background only makes the boundary conditions more
involved. It is not difficult to add the gauge field background as
in the model \cite{Hosotani:2006-2} to the above setup. When we
further reduce the 5-dimension spacetime to 4-dimension spacetime,
we can advance with the help of the model \cite{Hosotani:2006-2}.
$\lambda_n$ in the above setup corresponds to the bulk mass
parameter in 5-dimension spacetime in the model
\cite{Hosotani:2006-2}. Of course, it is obvious that one bulk mass
parameter corresponds to one family in 5-dimensions. Now we can
understand how the same order bulk mass parameters or families are
generated from extra space dimensions. The bulk mass parameters are
eigenvalues of Schr\"{o}dinger-like equations
(\ref{solution17})-(\ref{solution18}), so generally they should be
of the same order. The eigenstates of equations
(\ref{solution17})-(\ref{solution18}), belonging to the eigenvalues
$\lambda_n$, correspond to the generations in 5-dimensions. When we
reduce further the 5-dimension spacetime to the physical
4-dimensions, these generations in 5-dimensions can produce the
generations in 4-dimension spacetime. So the families in physical
4-dimension spacetime are generated simultaneously. Of course, the
eigenstates should be normalizable in order that we can get the
effective 5-dimension action after integrating out the sixth
dimension. We will discuss the normalization conditions in the next
section. However, a problem arises immediately from the following
contradiction: On one side, the eigenvalue problem of
Eqs.~(\ref{solution17})-(\ref{solution18}) is of the Sturm-Liouville
type. The characters of the Sturm-Liouville eigenvalue problem are
that the number of eigenvalues is infinite and the size of
eigenvalues is non-bounded, i.e., the eigenvalue series becomes
large monotonously; on the other side, as it has been illuminated
obviously in the papers \cite{Hosotani:2006-2}, the larger bulk mass
parameters produce lighter fermions mass in 4-dimensions. Therefore,
the eigenstates of Eqs.~(\ref{solution17})-(\ref{solution18})
produce infinite light fermion generations. However, no lighter
generations are discovered by experiments so far. Hence we need
another mechanics to cut off the infinite series and select only
several eigenstates. The left eigenstates correspond to generations
in 4-dimensions.

There also exists a problem about the zero mode (\ref{ZeroMode}). By
the numerical examples in \cite{Hosotani:2006-2}, the zero bulk mass
parameter in 5-dimensions produces a very heavy fermion in
4-dimensions, and it is heavier than the SM generations. So it does
not correspond to the physical generations. If the zero mode is
permitted by the boundary conditions and the normalization
conditions in our model, there would exist a generation that has not
been discovered by experiment so far. However, the zero mode is more
subtle in the example we will discussed in the next section. We will
discuss this problem in more detail in that section.

Now we suggest several approaches to deal with the problem about the
infinite eigenvalues.

(1) An immediate proposal is that one chooses a 6-dimension
spacetime in which the sixth dimension is not continuous but
discrete. For example, if we discrete the finite interval to be
finite points, the induced
Eqs.~(\ref{solution17})-(\ref{solution18}) will be difference
equations. The number of their eigenvalues is finite naturally.
There has been a similar investigation for gravity, see
\cite{Deffayet:2004}.

(2) Another bizarre proposal is that we suppose the sixth dimension
to be timelike. In the Ansatz Eq.~(\ref{ansatz1}), we have chosen
the sixth dimension to be spacelike. Instead we can choose it to be
timelike. This leads to the metric with two time dimensions
\cite{Overduin:1997}.\footnote{For extensive investigations on
two-time physics, see \cite{Bars:2006}} The alternative metric
Ansatz is
\begin{equation}
\label{metric2} ds^2 = B(z)^2\left[A(y)^2\left( \eta_{\mu \nu}
dx^{\mu} dx^{\nu} + d y^2\right)- d z^2\right].
\end{equation}
In Eq.~(\ref{Eq3}), we let $\Gamma^6 = \left(
\begin{array}{cc}
 0 & -\mathbf{1}_4 \\
\mathbf{1}_4 & 0 \end{array}\right)$, with others keeping invariant.
The same procedure produces the equations,
\begin{eqnarray}
\label{solution20} \frac{d^2}{d z^2}F_n(z)-B^{-1}\dot{B}\frac{d}{d
z}F_n(z)+m^2 F_n(z)
+\left[-\lambda_n B^{-1}\dot{B}-\lambda_n^2\right]F_n(z)=0,\\
\label{solution21} \frac{d^2}{d z^2}G_n(z)-B^{-1}\dot{B}\frac{d}{d
z}G_n(z)+m^2 G_n(z) +\left[\lambda_n
B^{-1}\dot{B}-\lambda_n^2\right]G_n(z)=0.
\end{eqnarray}
We give a simple example in which $B(z)=\mathrm{constant}$. The
solutions are
\begin{eqnarray}
\label{solution22} F(z)=C_1~ \mathrm{exp}^{ikz}+C_2~
\mathrm{exp}^{-ikz},~~k=\sqrt{m^2-\lambda^2}.
\end{eqnarray}
Here we have omitted the subscripts. If we impose the boundary
conditions,
\begin{eqnarray}
\label{solution23} F|_{0}=0,~~F|_{R}=0,
\end{eqnarray}
then $\lambda$ must conform to
\begin{eqnarray}
\label{solution24}
kR=\sqrt{m^2-\lambda^2}R=n\pi, n=1,2,3,\cdots.
\end{eqnarray}
As we emphasized above, $\lambda$ and $m$ must both be real numbers.
Eq. (\ref{solution24}) has solutions only for finite natural number.
The number of the eigenvalues depends on the size of $m$, hence
there are a finite number of eigenvalues.

(3) However, in the following section we will adopt another
approach, i.e., we can obtain a finite number of eigenstates by
choosing the metric $B(z)$ delicately. Although $B(z)$ produces
infinite eigenstates generally, it is possible that some of them can
produce only finite eigenstates. We will focus on this possibility
in the following section, and give concrete examples for finite
generations.

\section{Examples of metric for finite generations}\label{sec:3}

In this section, we will suggest metric examples which can produce
finite generations. We discuss the appropriate normalization
conditions for $F_{n}(z),~G_{n}(z)$ at first, then we analyze
examples in detail.

We begin with the action (\ref{action2}). With the help of Eqs.
(\ref{solution9}), (\ref{solution10}) and (\ref{solution11}), the
action (\ref{action2}) can be rewritten as
\begin{eqnarray}
\label{norm1} S &=&\int d^4 x dy~K_{mn}\left\{\frac{i}{2}
A^4\left[\bar{\psi}_m\gamma^5\partial_5\psi_n-\partial_5\bar{\psi}_m\gamma^5\psi_n+\bar{\psi}_m\gamma^{\mu}\partial_{\mu}\psi_n-\partial_{\mu}\bar{\psi}_m\gamma^{\mu}\psi_n\right]\right\}\nonumber\\
&-&\int d^4 x dy~M_{mn}A^5 i\bar{\psi}_m {\psi}_n, \\
\label{norm2} K_{mn}&=& \int dz B^5
\left(\widehat{F}_{m}^{\ast}\widehat{F}_n+\widehat{G}_{m}^{\ast}\widehat{G}_n\right)=\int
dz\left({F}_{m}^{\ast}{F}_n+{G}_{m}^{\ast}{G}_n \right),\\
\label{norm3} M_{mn}&=&\int dz B^5\left[
\left(\widehat{F}_{m}^{\ast}\widehat{F}_n+\widehat{G}_{m}^{\ast}\widehat{G}_n
\right)\frac{\lambda_n+\lambda_{m}^{\ast}}{2}\right]=\int dz
\left[\left({F}_{m}^{\ast}{F}_n+{G}_{m}^{\ast}{G}_n\right)\frac{\lambda_n+\lambda_{m}^{\ast}}{2}\right].
\end{eqnarray}
Notice that
$\bar{\Psi}=\Psi^{\dag}\Gamma^{0}=(\bar{\chi}_2,\bar{\chi}_1)$.
Eqs.~(\ref{norm1}-\ref{norm3}) are satisfied for all modes,
including zero modes and all massive modes.

In order to get the conventional effective 5-dimensional action
\begin{eqnarray}
\label{action3} S_{5eff}&=&\sum_n\int d^4 x dy\left\{\frac{i}{2}
A^4\left[\bar{\psi}_n\gamma^5\partial_5\psi_n-\partial_5\bar{\psi}_n\gamma^5\psi_n+\bar{\psi}_n\gamma^{\mu}\partial_{\mu}\psi_n-\partial_{\mu}\bar{\psi}_n\gamma^{\mu}\psi_n\right]\right\}\nonumber\\
&-&\sum_n \int d^4 x dy A^5 i\lambda_{n}\bar{\psi}_n {\psi}_n,
\end{eqnarray}
we consider two cases below:

Case~(I): The first case is that the normalization conditions
\begin{eqnarray}
\label{norm2a} K_{mn}=\int dz
\left({F}_{m}^{\ast}{F}_n+{G}_{m}^{\ast}{G}_n\right)=\delta_{mn}
\end{eqnarray}
are satisfied. As in the standard Sturm-Liouville case, we can
convert the normalization conditions (\ref{norm2a}) to the boundary
conditions. By Eqs.~(\ref{solution15}) and (\ref{solution16}), we
have
\begin{eqnarray}
\label{norm2b}
\left[\lambda_n^2-(\lambda_m^2)^{\ast}\right]\int_{R}^{R'}&dz&
\left({F}_{m}^{\ast}{F}_n+{G}_{m}^{\ast}{G}_n\right)\nonumber\\
&=&\left[\left(F_n\frac{d}{dz}F_m^{\ast}-F_m^{\ast}\frac{d}{dz}F_n\right)+\left(G_n\frac{d}{dz}G_m^{\ast}-G_m^{\ast}\frac{d}{dz}G_n\right)\right]|_{R}^{R'}\\
\label{norm2b-0}
&=&\left(\lambda_n+\lambda_m^{\ast})(F_m^{\ast}G_n-G_m^{\ast}F_n\right)|_{R}^{R'}.
\end{eqnarray}
In the last line, we have used the bulk equations (\ref{solution13})
and (\ref{solution14}) to simplify these expressions. We can also
get (\ref{norm2b-0}) from  Eqs.~(\ref{solution13}) and
(\ref{solution14}) directly. There are two types of concise choices
to make the normalization conditions satisfied,
\begin{eqnarray}
\label{norm2b-1} (a):~~F|_{R}&=&0,~F|_{R'}=0;\\
\label{norm2b-2}
\mathrm{or}~~(b):~~F|_{R}&=&G|_{R},~F|_{R'}=G|_{R'}.
\end{eqnarray}
Then for real $\lambda_m$, the orthogonality is ensured by
appropriate boundary conditions. In this case, we can get
Eq.~(\ref{action3}) from Eq.~(\ref{norm1}) via Eq.~(\ref{norm2a}) in
a straight way.

Case~(II): The second case is that the normalization conditions
Eq.~(\ref{norm2a}) are not satisfied. In this case, $K$ and $M$ are
both matrices, which means that different KK modes are mixed not
just among the mass terms, but also among the kinetic terms. At the
first sight, it seems that we can not get the the conventional
effective 5-dimensional action Eq.~(\ref{action3}). However, if $K$
is positive-definite and the number of KK modes is
finite\footnote{We will give numerical examples to show that such
conditions can be satisfied in Appendix \ref{sec:B}.}, we can
redefine the fermion field to get an action, which has the same form
with that of Eq.~(\ref{action3}). The difference is that the
eigenvalues $\lambda_n$ are modified to different size. From
Eq.~(\ref{norm2}) and Eq.~(\ref{norm3}), we know that $K$ and $M$
are both hermitian. A positive-definite hermitian matrix $K$ can be
diagonalized as
\begin{eqnarray}
\label{diag1} K&=&V^{\dagger}\Lambda V=H^{\dagger}H,~~H=\sqrt{\Lambda} V,\\
\Lambda&=&\mathrm{diag}\left(\Lambda_1,\Lambda_2,\cdots,\Lambda_n\right),\nonumber\\
\sqrt{\Lambda}&=&\mathrm{diag}(\sqrt{\Lambda_1},\sqrt{\Lambda_2},\cdots,\sqrt{\Lambda_n}).\nonumber
\end{eqnarray}
In the above expressions, $\Lambda_i>0,~i=1,2,\cdots,n$, as we have
supposed that $K$ is positive-definite. Redefine $\psi_n$ as
\begin{eqnarray}
\label{diag4} \widetilde{\psi}_m=H_{mn}\psi_n,
\end{eqnarray}
then in the new basis $\widetilde{\psi}_n$, $M$ becomes
\begin{eqnarray}
\label{diag5} \widetilde{M}=(H^{-1})^{\dagger}M H^{-1}.
\end{eqnarray}
After diagonalizing $\widetilde{M}$ by $U$, we have
\begin{eqnarray}
\label{diag6} \widetilde{M}&=&U^{\dagger}\Delta U,\\
\Delta&=&\mathrm{diag}(\widehat{\lambda}_1,\widehat{\lambda}_2,\cdots,\widehat{\lambda}_n).\nonumber
\end{eqnarray}
The action (\ref{norm1}) can be reduced to the form like that of
action (\ref{action3})
\begin{eqnarray}
\label{action3a} \widehat{S}_{5eff}&=&\sum_n\int d^4 x
dy\left\{\frac{i}{2}
A^4\left[\bar{\widehat{\psi}}_n\gamma^5\partial_5\widehat{\psi}_n-\partial_5\bar{\widehat{\psi}}_n\gamma^5\widehat{\psi}_n+\bar{\widehat{\psi}}_n\gamma^{\mu}\partial_{\mu}\widehat{\psi}_n-\partial_{\mu}\bar{\widehat{\psi}}_n\gamma^{\mu}\widehat{\psi}_n\right]\right\}\nonumber\\
&-&\sum_n \int d^4 x dy A^5
i\widehat{\lambda}_{n}\bar{\widehat{\psi}}_n {\widehat{\psi}}_n,\\
\widehat{\psi}_m&=&U_{mn}\widetilde{\psi}_n.\nonumber
\end{eqnarray}

In the above, we have given the normalization conditions. The
criteria are that we can integrate out the sixth dimension to get an
effective 5-dimension action. Now we suggest an example that can
produce finite generations. In Eqs.~(\ref{solution15}) and
(\ref{solution16}), we suppose that
\begin{eqnarray}
\label{suppose1} B(z)=s \frac{e^{\omega z}+a}{e^{\omega
z}+b},~~s,a,b,\omega>0.
\end{eqnarray}
As in the models \cite{Randall:1999}, $\omega$ can be regarded as
the characteristic energy scale of the sixth dimension. We will see
that it determines the size of KK modes below. The role of the
dimensionless parameters $s$ and $b$ will become obvious after we
give the solutions of Eqs.~(\ref{solution15}) and
(\ref{solution16}). The conditions $a,b>0$ ensure that the metric is
well behaved in the interval $(-\infty,\infty)$.

Eqs.~(\ref{solution15}) and (\ref{solution16}) can be solved by
hypergeometrical functions,
\begin{eqnarray}
\label{suppose1-2} F(z)&=&C_1 e^{-\mu \omega z}(e^{\omega
z}+b)^{\mu-\nu}
\mathrm{hypergeom}\left(\rho-\mu+\nu,1-\rho-\mu+\nu;1-2\mu,\frac{e^{\omega
z}}{e^{\omega z}+b}\right) \nonumber \\&+&C_2 e^{\mu \omega
z}(e^{\omega z}+b)^{-\mu-\nu}
\mathrm{hypergeom}\left(\rho+\mu+\nu,1-\rho+\mu+\nu;1+2\mu,\frac{e^{\omega
z}}{e^{\omega z}+b}\right),~~
\end{eqnarray}
where $\rho=\frac{m}{\omega}s(1-\frac{a}{b})$,
$\mu=\sqrt{\left(\frac{m}{\omega}s\right)^2\left(\frac{a}{b}\right)^2-\left(\frac{\lambda}{\omega}\right)^2}$,
and
$\nu=\sqrt{\left(\frac{m}{\omega}s\right)^2-\left(\frac{\lambda}{\omega}\right)^2}$.
$C_1$ and $C_2$ are constants. For the sake of simplicity, we omit
the subscript $n$. We only display the solution for $F(z)$
explicitly. The solution for $G(z)$ can be determined by $F(z)$
through Eq.~(\ref{solution13}) or by Eq.~(\ref{solution16})
directly. Now let us investigate this solution. For hypergeometrical
function\footnote{About the property of hypergeometrical function,
see Appendix
\ref{sec:C}.}$~\mathrm{Hypergeom}(\alpha,\beta;\gamma,\xi)$, when
$\mathrm{Re}(\gamma-\alpha-\beta)\leq 0$, it diverges at $\xi=1$. In
the solution (\ref{suppose1-2}), we have
\begin{eqnarray}
\label{supposea}
\mathrm{Re}(\gamma-\alpha-\beta)=\mathrm{Re}(-2\nu)\leq 0,
\end{eqnarray}
so when $z\rightarrow\infty$, $\xi=\frac{e^{\omega z}}{e^{\omega
z}+b}\rightarrow 1$, a singularity happens. If we choose $z$ to be a
finite interval $[R,R']$, then the solution (\ref{suppose1-2}) is
well behaved in this range. Imposing the boundary conditions
(\ref{norm2b-1}) or (\ref{norm2b-2}), we get infinite eigenvalues
generally. However, we find that the following choice can produce a
finite number of eigenvalues. Given that $z$ to be a semi-infinite
interval $[R,\infty)$, then the solution (\ref{suppose1-2}) develops
a singularity when $z\rightarrow\infty$. This singularity makes the
integral in (\ref{norm2a}) to be divergent. In order to make the
integral to be finite, the hypergeometrical series
(\ref{suppose1-2}) must be cut off to be a polynomial by the
requirement $\alpha=-(n-1),\mathrm{or}~\beta=-(n-1),n=1,2,3,\cdots$.
In Eq.~(\ref{suppose1-2}), we choose
\begin{eqnarray}
\label{suppose1-5} 1-\rho-\mu+\nu=-(n-1),~n=1,2,3,\cdots .
\end{eqnarray}
Because $\lambda$ and $m$ are real numbers, as we have discussed
above, Eq.~(\ref{suppose1-5}) might have solutions if
\begin{eqnarray}
\label{con1} 1\leq n \leq
\left[\sqrt{\left(\frac{a}{b}\right)^2-1}-\left(\frac{a}{b}-1\right)\right]\frac{m}{\omega}s.
\end{eqnarray}
Obviously, the size of $n$ are limited by the parameters in the
metric, then only finite eigenvalues are permitted. We see that it
is important that $\lambda$ and $m$ are real numbers again. The
conditions (\ref{suppose1-5}) are required by the boundary
conditions when $z\rightarrow\infty$. This boundary condition
restricts the solutions to the form
(\ref{Massive-1})-(\ref{Massive-3}). We display these solutions in
Appendix \ref{sec:A} explicitly. We see that these solutions are
determined completely up to normalization constants. For these
solutions, $F_n(z),~G_n(z)\rightarrow x^{-\nu_n}$ in the symbols in
Appendix \ref{sec:A}, when $z\rightarrow\infty$. The integral in
(\ref{norm2a}) is well defined in the interval $[R,\infty)$, if
$\nu_n>0,x>0$. Because we have chosen the range of $z$ as the
interval $[R,\infty)$, we should also discuss the boundary
conditions at $z=R$. We might want to follow the discussions in
Case~(I), that is, we require that the normalization conditions
(\ref{norm2a}) are satisfied. Because $F_n(z),~G_n(z)\rightarrow 0$
when $z\rightarrow\infty$, the normalization conditions
(\ref{norm2a}) require that
\begin{eqnarray}
\label{norm2b-3}
\left[\lambda_n^2-(\lambda_m^2)^{\ast}\right]\int_{R}^{R'}&dz&
\left({F}_{m}^{\ast}{F}_n+{G}_{m}^{\ast}{G}_n\right)\nonumber\\
&=&-\left[\left(F_n\frac{d}{dz}F_m^{\ast}-F_m^{\ast}\frac{d}{dz}F_n\right)+\left(G_n\frac{d}{dz}G_m^{\ast}-G_m^{\ast}\frac{d}{dz}G_n\right)\right]|_{R}\nonumber\\
&=&-\left(\lambda_n+\lambda_m^{\ast})(F_m^{\ast}G_n-G_m^{\ast}F_n\right)|_{R}=0.
\end{eqnarray}
However, it is difficult to require the solutions
(\ref{Massive-1})-(\ref{Massive-3}) to satisfy the conditions
(\ref{norm2b-3}). From (\ref{suppose1-5}), we know that $\lambda_n$
are determined by the parameters $\frac{m}{\omega}s,\frac{a}{b}$ and
$n$. So the conditions (\ref{norm2b-3}) impose restrictions on the
parameters $\frac{m}{\omega}s,\frac{a}{b}$ and the boundary
parameter $R$ instead of $\lambda_n$. The naive numerating of
parameters may mean that we can have 3 eigenstates to be orthogonal,
because we have 3 parameters $\frac{m}{\omega}s,\frac{a}{b}$ and
$R$. Nevertheless, such choices are difficult to be implemented and
it seems less natural. It is more natural to regard the parameters
$\frac{m}{\omega}s,\frac{a}{b}$ and $R$ as the input parameters, or
they should be determined by unknown physics that we do not consider
here. In the following discussions, we will not impose boundary
conditions further at $z=R$ to determine the solutions, but we will
simply give these parameters by hand to determine the solutions. In
such a choice, the normalization conditions (\ref{norm2a}) are not
satisfied. So we should change to the Case~(II), that is, $K$ is
matrix valued, and we try to diagonalize this matrix to get the
action (\ref{action3a}). Before doing that, we give numerical
examples to show that only three eigenvalues are left and they are
of the same order. Let $\frac{a}{b}=9/4,\frac{m}{\omega}s=4$. From
Eq.~(\ref{con1}), we know that only $n=1,2,3$ are permitted. The
eigenvalues are given by
\begin{eqnarray}
\label{numerical-1} \lambda_1
=3.187~\omega~(n=1),~\lambda_2=3.833~\omega~(n=2),~\lambda_3=3.9995~\omega~(n=3),
\end{eqnarray}
which are of the same order. These massive modes together with the
zero mode make $K$ and $M$ to be $4\times 4$ matrices. In order to
get the action (\ref{action3a}), $K$ must be positive-definite. In
Appendix \ref{sec:B}, we give numerical examples to show that $K$ is
positive-definite and the modified eigenvalues $\widehat{\lambda}_n$
in (\ref{diag6}) are still of the same order. $\widehat{\lambda}_n$
are given by
\begin{eqnarray}
\label{numerical-2} \widehat{\lambda}_0
=-2.69625~\omega,~\widehat{\lambda}_1=4.00816~\omega,~\widehat{\lambda}_2=4.44389~\omega,
~\widehat{\lambda}_3=5.26792~\omega.
\end{eqnarray}

Here we give some interpretations for our choices of the parameters.
$\frac{a}{b}=9/4,\frac{m}{\omega}s=4$ are chosen to ensure that only
3 massive modes are permitted. We choose $x_0=\frac{e^{\omega
R}}{b}=30$ as the boundary value in order to ensure that the
modified $\widehat{\lambda}_n$ are still of the same order. We found
that small $x_0$, for example, $x_0=1$, makes different
$\widehat{\lambda}_n$ to have big difference. Now we can understand
the role of the parameters $s$ and $b$ in the metric
(\ref{suppose1}). In order that we can trust our analysis, the
condition $\frac{m}{\omega}<1$ should be satisfied. From the above,
we see that $s$ appears in the combination $\frac{m}{\omega}s$. Then
we can always keep $\frac{m}{\omega}<1$ by adjusting the value of
$s$ despite of the input value of $\frac{m}{\omega}s$. While $b$
appears in the combination $\frac{e^{\omega z}}{b}$, so it is
closely related to the boundary value of $z$. The role of $a$ is
less obvious because it appears in a more complex way.

From the above, we notice two obvious changes: (1) The massive modes
are modified to different size, but they are still of the same
order; (2) The zero mode mixes with the massive modes. By this
mixing, the zero mode gets mass of the same order with the massive
modes. A strange point is that the zero mode gets a negative mass.
However, it does not form problems in the models like
\cite{Hosotani:2006-2}, where only the size of the mass is relevant.
In models where the sign of mass is relevant, we must reconsider
whether it produces problems for our model. This new feature can
supply a possibility to bypass the zero mode problem that we
introduced in Sec.~\ref{sec:2}. In the above, we get 4 massive modes
from the previous 3 massive modes and 1 zero mode. They can produce
4 fermion generations in 4-dimension. This is not realistic. The
above numerical example suggests us to start with 2 massive modes
and 1 zero mode. If the zero mode gets mass of the same order with
the massive modes through mixing as the above numerical example, we
can get only 3 generations.

Before giving an example about this situation, we should discuss
another mass source about the zero mode, that is, the zero mode can
also get mass through coupling with a Higgs field on the brane sited
at $z=R$. Here it is appropriate to introduce the brane coupling.
The Wilson line phase in \cite{Hosotani:2006-2} is not well defined
because the range of $z$ is noncompact. We do not suggest a concrete
form for this coupling. For example, it can arise from the coupling
used in \cite{Kaplan:2001}. Here we accept the result that the zero
mode gets mass
\begin{eqnarray}
\label{ZeroM} \lambda_0=\epsilon~\omega.
\end{eqnarray}
In the following discussions, we will consider two cases: (a)
$\epsilon\rightarrow 0$, so it is negligible; (b) $\epsilon\sim 1$,
so it is comparable with the massive modes. We will give numerical
examples about these two cases respectively. According to the same
spirit with the above example, we let
$\frac{a}{b}=8/3,\frac{m}{\omega}s=3$ to ensure that only 2 massive
modes are permitted. These massive modes are given by
\begin{eqnarray}
\label{numerica3} \lambda_1
=2.548~\omega~(n=1),~\lambda_2=2.969~\omega~(n=2).
\end{eqnarray}
We still choose $x_0=\frac{e^{\omega R}}{b}=30$, then by the same
procedure with that in Appendix \ref{sec:B}, we can get the modified
$\widehat{\lambda}_n$ as
\begin{eqnarray}
\label{numerical4} \widehat{\lambda}_0
&=&-1.48838~\omega,~\widehat{\lambda}_1=3.16287~\omega,~\widehat{\lambda}_2=3.85202~\omega,
~\mathrm{for}~\epsilon=0;\\
\label{numerical5} \widehat{\lambda}_0
&=&1.71708~\omega,~\widehat{\lambda}_1=2.59758~\omega,~\widehat{\lambda}_2=3.19837~\omega,
~\mathrm{for}~\epsilon=2.
\end{eqnarray}
Here we only give the results. The details are similar to that in
Appendix \ref{sec:B}. The zero mode gets mass of the same order with
the massive modes in both cases. The difference is that the sign of
the zero mode mass is opposite in these two cases. In both cases,
with the help of the zero mode mixing with massive modes or the zero
mode coupling with Higgs field, we may suggest a possibility to
bypass the zero mode problem in Sec.~\ref{sec:2}. In each case, we
may get just 3 generations. In models where the sign of mass is
relevant, we can check which case may be realistic.

In the above example, the orthogonal conditions (\ref{norm2a}) are
not satisfied, because we choose a special metric and a special
range for $z$. This choice induces mixing between different modes.
In the following, we try to construct an example, which can ensure
that the orthogonal conditions (\ref{norm2a}) are satisfied. This
example can be constructed by changing the metric (\ref{suppose1})
to the following form,
\begin{eqnarray}
\label{suppose2} B(z)=s \frac{e^{\omega z}-a}{e^{\omega
z}-b},~~s,a,b,\omega
>0.
\end{eqnarray}
Because $a,b>0$, this metric develops singularity\footnote{See
Appendix \ref{sec:F}.} at the point
$z=\frac{\mathrm{log}a}{\omega}$. This may make this metric
unrealistic. However, we find that it can satisfy the orthogonal
conditions (\ref{norm2a}) just because it has such special
structure. Here we let aside the problem of singularity, and focus
on how it can satisfy the orthogonal conditions. The solutions can
be given by hypergeometrical functions yet,
\begin{eqnarray}
\label{suppose2-2} \widehat{F}(z)&=&C_1 e^{-\mu \omega z}(e^{\omega
z}-b)^{\rho}
\mathrm{hypergeom}\left(\rho-\mu-\nu,\rho-\mu+\nu;1-2\mu,\frac{e^{\omega
z}}{b}\right)\\ \nonumber &+&C_2 e^{\mu \omega z}(e^{\omega
z}-b)^{\rho}
\mathrm{hypergeom}\left(\rho+\mu-\nu,\rho+\mu+\nu;1+2\mu,\frac{e^{\omega
z}}{b}\right),
\end{eqnarray}
where $\rho,\mu$ and $\nu$ keep the same form with that in
Eqs.~(\ref{suppose1-2}). By use of the property of the
hypergeometrical function, we see that if $\mathrm{Re}(1-2\rho)\leq
0$, a singularity happens at $z=R=\frac{\mathrm{log}b}{\omega}$.
Then if we choose the range of $z$ to be $(-\infty,~R]$, like the
above example, the boundary conditions at
$z=R=\frac{\mathrm{log}b}{\omega}$ impose the conditions like
(\ref{suppose1-5}). We should also make the solutions well behaved
when $z\rightarrow-\infty$. These two requirements can be satisfied
by the following condition
\begin{eqnarray}
\label{suppose2-4} \rho+\mu-\nu=-n,~n=0,1,2,\cdots .
\end{eqnarray}
For $\frac{a}{b}<1$, Eq.~(\ref{suppose2-4}) has solutions and $n$ is
limited by
\begin{eqnarray}
\label{suppose2-5} 0\leq n\leq
\left[\sqrt{1-\left(\frac{a}{b}\right)^2}-\left(1-\frac{a}{b}\right)\right]\frac{m}{\omega}s.
\end{eqnarray}
We give these solutions in Appendix \ref{sec:D} explicitly. We
suppose that $\frac{m}{\omega}s=10,~\frac{a}{b}=\frac{1}{2}$, in
order that only 3 massive modes are permitted. Besides these massive
modes, Eq.~(\ref{suppose2-4}) also has a zero mode solution. These
solutions are given by
\begin{eqnarray}
\label{suppose2-6} \lambda_0 =0,~\lambda_1=3.80~\omega,~\lambda_2
=4.642~\omega,~\lambda_3=4.953~\omega.
\end{eqnarray}
From Appendix \ref{sec:D}, we know that when $z\rightarrow-\infty$,
then $x\rightarrow 0$, $F_n(z),~G_n(z)\rightarrow 0$; while at
$z=R=\frac{\mathrm{log}b}{\omega}$, $x=1$, then $F_n(z),~G_n(z)=0$.
So in this example, the boundary conditions (\ref{norm2b-1}) are
satisfied. Then the orthogonal conditions (\ref{norm2a}) are
ensured. There are no mixing among different modes. So in this
example, the zero mode can only become massive through coupling with
Higgs field. If the zero mode can get mass comparable to the massive
modes, we can adjust the parameters $\frac{m}{\omega}s$ and
$\frac{a}{b}$ to make that only 2 massive modes are permitted. Then
we can get just 3 generations. If this zero mode gets small mass, a
generation heavier than the SM generations is produced. We should
check whether it is allowed by experiments. If it is objected by
experiments, it will make a problem for our model.

In Appendix \ref{sec:E}, we suggest another example, in which the
orthogonal conditions (\ref{norm2a}) are satisfied. Like the metric
(\ref{suppose2}), there exists a singularity in the range of $z$ we
considered. In order to avoid the singularity, we may change the
range of $z$. For example, for the metric (\ref{suppose2}), we can
choose the range of $z$ to be $[z_1,R]$, where
$z_1>\frac{\mathrm{log}a}{\omega}$. In this new range, the
singularity of the metric (\ref{suppose2}) at
$\frac{\mathrm{log}a}{\omega}$ is avoided. But the orthogonal
conditions (\ref{norm2a}) will be not satisfied. We have not found a
metric which satisfies the requirements: 1) it can produce finite
generations; 2) it ensures that the orthogonal conditions
(\ref{norm2a}) are satisfied, and is well behaved in the range of
$z$.

\section{Further Discussions}\label{sec:4}

In this section, we compare the approach adopted in
\cite{Libanov:2001,Nernov:2002} with our setup in Sec.~\ref{sec:2}
at first. In the  approach adopted in
\cite{Libanov:2001,Nernov:2002}, the authors reduced the 6-dimension
spacetime to 4-dimension spacetime directly. Distinct from this
approach, we reduce two layer warped 6-dimension (4+1)+1 spacetime
to the 5-dimension 4+1 spacetime which is still warped at the first
step, then we reduce the 5-dimension spacetime to the physical
4-dimension spacetime. Their differences are the different ways that
one treats the zero mode and massive modes.

(1) For the zero mode: In the approach adopted in
\cite{Libanov:2001,Nernov:2002}, the authors reduced the 6-dimension
spacetime to 4-dimensions directly, and got zero modes in
4-dimension spacetime, so these zero modes correspond to the
standard model~(SM) generations. These zero modes get mass through
coupling to the Higgs field. In the present approach, we reduce the
6-dimension spacetime to 5-dimensions at first, so we get a zero
mode in 5-dimension spacetime. When one further reduces the
5-dimension spacetime to the physical 4-dimension spacetime, the
zero mode in 5-dimension spacetime can produce a very heavy fermion
in 4-dimensions, as illuminated obviously in \cite{Hosotani:2006-2}.
It is very heavy, hence it does not correspond to the SM generation.
However, as we discussed in Sec.~\ref{sec:3}, the zero mode can also
become massive, then it can produce SM generation if it can get
large mass. But if it gets small mass, then it produces a new
generation objected by experimental data. This can cause a problem
for our model.

(2) For the massive modes: In the approach adopted in
\cite{Libanov:2001,Nernov:2002}, the authors got the massive modes
in 4-dimension spacetime. The massive modes are heavy
Kluza-Klein~(KK) particles. They do not correspond to the SM
fermions. However, in the present approach, we reduce the
6-dimension spacetime to 5-dimensions at first, so we get the
massive modes in 5-dimension spacetime again. These massive modes
are KK states in 5-dimension spacetime. When one reduces further the
5-dimension spacetime to the physical 4-dimension spacetime, these
massive modes in 5-dimension spacetime can produce massive fermions
in 4-dimension spacetime, which are light and correspond to the SM
fermions, as it is obvious in \cite{Hosotani:2006-2}.

These differences provide a new chance that we can address some
extra issues in the present approach: we give an explanation for the
origin of the same order bulk mass parameters, and give an answer
for the fermion generation puzzle in the meantime. Note that it is
the special metric Ansatz (\ref{ansatz1}) to supply such an
explanation. The two layer structure of the metric enables one to
reduce the 6-dimension spacetime to 5-dimension spacetime. The
metric Ansatz for $A(y)$ can be the AdS metric,
\begin{eqnarray}
\label{ansatz5} A(y)=\frac{1}{k y}.
\end{eqnarray}
By this choice, the five dimensional Dirac equation
(\ref{solution9}) is similar to that analyzed in models
\cite{Hosotani:2006-2}.

Besides, we want to address another two issues:

(1) Whether the metric (\ref{ansatz1}) can be the background
solutions of Einstein equations? The metric Ansatz (\ref{ansatz1})
has been analyzed in \cite{Collins:2001} in high derivative gravity
with matter sources and in \cite{Choudhury:2007} with a negative
bulk cosmological constant. Their solutions are not the metric which
we suggested in Sec.~\ref{sec:3}. Here we consider a minimum coupled
scalar-gravity system in order to investigate whether the metric in
Sec.~\ref{sec:3} can be realized. This is a simple and convenient
way. The action is given by
\begin{eqnarray}
\label{einstein1} S&=&\int d^4 x dy dz \sqrt{-g} \{2M^4 R\} + \int
d^4 x dy dz \sqrt{-g}
\left\{\frac{1}{2}g^{MN}\nabla_{M}\phi\nabla_{N}\phi+V(\phi)\right\},
\end{eqnarray}
in which $V(\phi)$ is the potential term for scalar field. Supposing
that the metric Ansatz (\ref{ansatz1}) and that $\phi$ only depends
on $z$, we get the following equations,
\begin{eqnarray}
\label{einstein2}
4B^{-1}B_{zz}+2B^{-2}B^2_{z}+3A^{-3}A_{yy}&=&\frac{1}{4M^4}\left[-B^2\left(\frac{1}{2}B^{-2}
\phi^2_{z}+V(\phi)\right)\right],\\
\label{einstein3}
4B^{-1}B_{zz}+2B^{-2}B^2_{z}+6A^{-4}A^2_{y}&=&\frac{1}{4M^4}\left[-B^2\left(\frac{1}{2}B^{-2}
\phi^2_{z}+V(\phi)\right)\right],\\
\label{einstein4}
10B^{-2}B^2_{z}+4A^{-3}A_{yy}+2A^{-4}A^2_{y}&=&\frac{1}{4M^4}\left[\phi^2_{z}-B^2\left(\frac{1}{2}B^{-2}
\phi^2_{z}+V(\phi)\right)\right],\\
\label{einstein5} B^{-2}\phi_{zz}+4B^{-3}B_{z} \phi_z-\frac{d
V(\phi)}{d \phi}&=&0,
\end{eqnarray}
in which $A_y=\frac{dA}{dy},~B_z=\frac{dB}{dz},~\phi_z=\frac{d
\phi}{dz}$. Obviously, $A(y)$ should be of the form
$A(y)=\frac{1}{ky+c}$, in which $k,~c$ are constant. A minimum
coupled scalar-gravity coupled system has been analyzed in
\cite{DeWolfe:2000} in five dimensions. The result is that there
always exists appropriate form of $V(\phi)$ to ensure that the
metric have solutions, and $V(\phi)$ and the metric can be expressed
with a superpotential. The similar result applies to the above
system, that is, for any $B(z)$, there exists appropriate $V(\phi)$,
which makes Eqs. (\ref{einstein2})-(\ref{einstein5}) satisfied.
However, the present system is more complex, and it is difficult to
express the solutions with a superpotential. Of course, in order to
get the solution of the metric adopted in Sec.~\ref{sec:3}, the
boundary conditions must be adopted appropriately.

 (2) Whether the examples used in Sec.~\ref{sec:3} are
acceptable physically? The metric (\ref{suppose1}), (\ref{suppose2})
and (\ref{suppose3}) are similar to that analyzed in
\cite{Gregory:2001}, that is, they are both asymptotically flat.
They are also both noncompact, and both have infinite volume
\cite{Dvali:2001}. Especially for the metric (\ref{suppose2}) and
the metric (\ref{suppose3}), there is singularity
\cite{Gellmann:1985} in the range of $z$ that we choose. So it needs
further work to investigate whether they are acceptable physically.
The similar problem exists for the metric (\ref{metric2}), in which
the sixth space dimension is timelike. As being emphasized in
\cite{Dvali:1999}, the violations of casuality and probability give
stringent restrictions on the timelike dimension. It also needs
further work to investigate whether it is acceptable physically. We
have not found a metric, which is finite volumed, as in the models
\cite{Randall:1999}, and can produce finite generations
simultaneously.

\section{Conclusions}\label{sec:5}

Now we summarize the main points in our work. In this paper, we try
to explain the origin of the same order bulk mass parameters, and
give answers to the generation replication puzzle simultaneously.
The fermion masses are of hierarchy structure in 4-dimension
spacetime. It seems that it is difficult to interpret them as the
eigenvalues of a Schr\"{o}dinger-like equation. However, the
hierarchy structure can be reproduced with the bulk mass parameters
in 5-dimension spacetime. The 5-dimension mass parameters are in the
same order, as have been shown in many papers
\cite{Kaplan:2001,Csaki:2005,Hosotani:2006-1}. This interesting
feature supplies a chance to interpret the 5-dimension mass
parameters, which are of the same order, as the the eigenvalues of a
Schr\"{o}dinger-like equation. Supposing that the six dimension
spacetime metric has special two layer (4+1)+1 structure, we can
reduce the 6-dimension spacetime to 5-dimension spacetime at the
first step. We find that the bulk mass parameters are the
eigenvalues of a Schr\"{o}dinger-like equation. Hence the same order
mass parameters emerge naturally. However, the problem is that the
number of eigenvalues is infinite generally, which leads to infinite
light generations. We suggest several approaches to deal with this
problem. Obviously, this problem arises from the fact that in the
conventional Kluza-Klein~(KK) decomposition, one gets infinite KK
particles generally. However, as in the example given by Madore
\cite{Madore:1990}, in the noncommutative geometrical background,
and by the choice of the internal structure, the modification of KK
theory gives rise to finite spectrum of particles. Therefore, it is
possible to overcome the difficulties in our work by use of the
noncommutative geometry. It will require modifying the framework in
Sec.~\ref{sec:2}. We hope we can address these issues in the future.

{\bf Acknowledgement} This work is partially supported by National
Natural Science Foundation of China (Nos.~10721063, 10575003,
10528510), by the Key Grant Project of Chinese Ministry of Education
(No.~305001), by the Research Fund for the Doctoral Program of
Higher Education (China).

\appendix

\section{Explicit solutions for massive modes and zero mode: metric I}\label{sec:A}

In this appendix, we give the solutions of Eq.~(\ref{suppose1-2})
for massive modes under the conditions (\ref{suppose1-5})
explicitly. The solutions should be well behaved in the range
$[R,\infty)$. Let $x=\frac{e^{\omega z}}{b}$, the solutions are
given by
\begin{eqnarray}
\label{Massive-1}
F_1(x)&=&\frac{\sqrt{\omega}}{\sqrt{N_1}}x^{-\mu_1}(x+1)^{\mu_1-\nu_1},\\
\label{Massive-2}
F_2(x)&=&\frac{\sqrt{\omega}}{\sqrt{N_2}}x^{-\mu_2}(x+1)^{\mu_2-\nu_2}\left[1-\frac{\alpha_1}{\gamma_1}\frac{x}{x+1}\right],\\
\label{Massive-3}
F_3(x)&=&\frac{\sqrt{\omega}}{\sqrt{N_3}}x^{-\mu_3}(x+1)^{\mu_3-\nu_3}\left[1-\frac{2\alpha_3}{\gamma_3}\frac{x}{x+1}+\frac{\alpha_3(\alpha_3+1)}{\gamma_3(\gamma_3+1)}\left(\frac{x}{x+1}\right)^2\right],\\
\vdots  \nonumber
\end{eqnarray}
in which
\begin{eqnarray}
\label{Massive-1a}\gamma_n&=&1-2\mu_n,~~n=1,2,3,\cdots\\
\label{Massive-1b}\alpha_n&=&\rho-\mu_n+\nu_n,~\rho=\frac{m}{\omega}s\left(1-\frac{a}{b}\right),\\
\label{Massive-1c}\nu_n&=&\sqrt{\left(\frac{m}{\omega}s\right)^2-\left(\frac{\lambda_n}{\omega}\right)^2},\\
\label{Massive-1d}\mu_n&=&\sqrt{\left(\frac{m}{\omega}s\right)^2\left(\frac{a}{b}\right)^2-\left(\frac{\lambda_n}{\omega}\right)^2}.
\end{eqnarray}
We have dropped the hypergeometrical function after the coefficient
$C_2$ in (\ref{suppose1-2}), because it is divergent under the
conditions (\ref{suppose1-5}) when $z\rightarrow\infty$. The
solutions for $G_n(z)$ can be gotten from (\ref{solution13}) as
\begin{eqnarray}
\label{Massive-G}
G_{n}(x)=\frac{\omega}{\lambda_n}\left[\frac{m}{\omega}s\frac{x+\frac{a}{b}}{x+1}F_{n}(x)-x\frac{d}{dx}F_{n}(x)\right].
\end{eqnarray}
The zero mode solution is given by
\begin{eqnarray}
\label{ZeroMode1}
F_{0}(x)=0,~G_{0}(x)=\frac{\sqrt{\omega}}{\sqrt{N_0}}x^{-\frac{m}{\omega}s\frac{a}{b}}(x+1)^{\frac{m}{\omega}s(\frac{a}{b}-1)}.
\end{eqnarray}
When $z\rightarrow\infty$, $x=\frac{e^{\omega
z}}{b}\rightarrow\infty$
\begin{eqnarray}
\label{approx1} F_{n}(x)&\rightarrow&
x^{-\nu_n},~G_{n}(x)\rightarrow
\frac{\omega}{\lambda_n}\left(\frac{m}{\omega}s+\nu_n\right)x^{-\nu_n},\\
\label{approx2} F_{0}(x)&=&0,~G_0(x)\rightarrow
x^{-\frac{m}{\omega}s} .
\end{eqnarray}
They are all well behaved when $\nu_n>0$. In terms of $x$, the
integral in Eq.~(\ref{norm2}) can be rewritten as
\begin{eqnarray}
\label{AppB-2} K_{mn}=\int dz
\left({F}_{m}^{\ast}{F}_n+{G}_{m}^{\ast}{G}_n\right)=\frac{1}{\omega}\int
\frac{dx}{x} \left({F}_{m}^{\ast}{F}_n+{G}_{m}^{\ast}{G}_n\right).
\end{eqnarray}
They are also well behaved in the range $[R,\infty)$ when $\nu_n>0$
and $x>0$. In the numerical examples we give in Sec.~\ref{sec:3},
the conditions $\nu_n>0$ and $x>0$ are always satisfied.

\section{Numerical examples for finite generations}\label{sec:B}

In order to get numerical results, we need to input the parameters
$\frac{m}{\omega}s,\frac{a}{b}$ and $R$. From the solutions in
Appendix A, we know that it is enough to input value for
$x_0=\frac{e^{\omega R}}{b}$.

Let $\frac{m}{\omega}s=4,\frac{a}{b}=9/4$ as in Sec.~\ref{sec:3},
then for massive modes, only $ n=1,2,3$ are permitted. We further
designate $x_0=\frac{e^{\omega R}}{b}=30$. We normalize the
solutions for massive modes and zero mode according to
(\ref{norm2a}) as
\begin{eqnarray}
\label{norm2a-1} \int dz
\left({F}_{n}^{\ast}{F}_n+{G}_{n}^{\ast}{G}_n\right)=1,~n=0,1,2,3.
\end{eqnarray}
These conditions determine the normalization constants. Then the
matrix $K$ is determined to be
\begin{eqnarray}
\label{AppB-1} K=\left(
\begin{array}{cccc}
 1 & 0.8797 & 0.7144 & 0.3503 \\
 0.8797 & 1 & 0.9399 & 0.5173 \\
 0.7144 & 0.9399 & 1 & 0.6433 \\
 0.3503 & 0.5173 & 0.6433 & 1
\end{array}
\right) .
\end{eqnarray}
The indices of $K$ are determined according to (\ref{norm2}) as
\begin{eqnarray}
\label{AppB-2} K_{mn}=\int dz
\left({F}_{m}^{\ast}{F}_n+{G}_{m}^{\ast}{G}_n\right),~m,n=0,1,2,3.
\end{eqnarray}
$K$ can be diagonalized as
\begin{eqnarray}
\label{AppB-3} K&=&V^{T}\Lambda V,~\Lambda=\mathrm{diag}(3.0656,0.707478,0.211924,0.0149942), \\
V&=&\left(
\begin{array}{cccc}
 -0.491317 & -0.555244 & -0.544655 & -0.391998 \\
 -0.488159 & -0.224439 & 0.0638874 & 0.84098 \\
 -0.651525 & 0.221179 & 0.626183 & -0.366729 \\
 0.309555 & -0.769683 & 0.554224 & -0.0678292
\end{array}
\right),\nonumber
\end{eqnarray}
where $V^{T}$ means the transpose of $V$. This example shows that
$K$ is positive-definite, as we expected.

$\lambda_n$ is given by (\ref{numerical-1}) in Sec.~\ref{sec:3}. $M$
is determined by (\ref{norm3}) as
\begin{eqnarray}
\label{AppB-4} M_{mn}=\int dz
\left[\left({F}_{m}^{\ast}{F}_n+{G}_{m}^{\ast}{G}_n\right)\frac{\lambda_n+\lambda_{m}^{\ast}}{2}\right],
m,n=0,1,2,3.
\end{eqnarray}
Given the parameters, $M$ is determined to be
\begin{eqnarray}
\label{AppB-5} M=\left(
\begin{array}{cccc}
 0 & 1.402 & 1.3693 & 0.7006 \\
 1.402 & \frac{\sqrt{1463}}{12} & 3.2993 & 1.8589 \\
 1.3693 & 3.2993 & \frac{12 \sqrt{5}}{7} & 2.5196 \\
 0.7006 & 1.8589 & 2.5196 & \frac{3 \sqrt{455}}{16}
\end{array}
\right) .
\end{eqnarray}
Following the procedure in Sec.~\ref{sec:3}, we can get
$\widetilde{M}$ as
\begin{eqnarray}
\label{AppB-5}\widetilde{M}&=&\left(
\begin{array}{cccc}
 2.73442 & -1.35109 & -2.28483 & 2.23697 \\
 -1.35109 & 3.00477 & -1.47008 & 1.60816 \\
 -2.28483 & -1.47008 & 2.19631 & 1.7851 \\
 2.23697 & 1.60816 & 1.7851 & 3.08822
\end{array}
\right)=U^{T}\Delta U,\nonumber \\
U&=&\left(
\begin{array}{cccc}
 0.671378 & 0.216423 & -0.209452 & 0.677158 \\
 -0.442926 & 0.350635 & 0.637256 & 0.524191 \\
 0.256197 & -0.814717 & 0.495127 & 0.159525 \\
 -0.536126 & -0.407986 & -0.552162 & 0.491155
\end{array}
\right),\\
\Delta&=&\mathrm{diag}(5.26792,4.44389,4.00816,-2.69625).\nonumber
\end{eqnarray}

\section{Property of hypergeometrical function}\label{sec:C}

We cite a theorem \cite{Andrews:2000} about the hypergeometrical
function $F(\alpha,~\beta;~\gamma,\xi)$.

Theorem $2.1.2$  ~~~The series $F(\alpha,~\beta;~\gamma,\xi)$ with
$|\xi|=1$ converges absolutely if
$\mathrm{Re}(\gamma-\alpha-\beta)>0$. The series converges
conditionally if $\xi=e^{i\theta}\neq 1$ and $0\geq
\mathrm{Re}(\gamma-\alpha-\beta)>-1$ and the series diverges if
$\mathrm{Re}(\gamma-\alpha-\beta)\leq -1$.

\section{Solutions for zero mode and massive modes: metric II}\label{sec:D}

For the metric (\ref{suppose2}), the solutions for massive modes
well behaved in the range $(-\infty,R]$ are given by
\begin{eqnarray}
\label{AppD-1}
F_1(x)&=&\frac{\sqrt{\omega}}{\sqrt{N_1}}x^{\mu_1}(1-x)^{\rho}\left[1-\frac{\beta_1}{\gamma_1}x\right],\\
\label{AppD-2}
F_2(x)&=&\frac{\sqrt{\omega}}{\sqrt{N_2}}x^{\mu_2}(1-x)^{\rho}\left[1-\frac{2\beta_2}{\gamma_2}x+\frac{\beta_2(\beta_2+1)}{\gamma_2(\gamma_2+1)}x^2\right],\\
\vdots\nonumber
\end{eqnarray}
in which
\begin{eqnarray}
\label{AppD-1a}\gamma_n&=&1+2\mu_n,~~n=1,2,3,\cdots\\
\label{AppD-1b}\beta_n&=&\rho+\mu_n+\nu_n,\\
\label{AppD-1c}\nu_n&=&\sqrt{\left(\frac{m}{\omega}s\right)^2-\left(\frac{\lambda_n}{\omega}\right)^2},\\
\label{AppD-1d}\mu_n&=&\sqrt{\left(\frac{m}{\omega}s\right)^2\left(\frac{a}{b}\right)^2-\left(\frac{\lambda_n}{\omega}\right)^2}.
\end{eqnarray}
In the above expressions, we have defined $x=\frac{e^{\omega
z}}{b}$. $\rho$ is defined as in Sec.~\ref{sec:3}. $\lambda_n$ are
determined by (\ref{suppose2-4}), and $n$ is limited by
(\ref{suppose2-5}).  We have dropped the hypergeometrical function
after the coefficient $C_1$ in (\ref{suppose2-2}), because it is
divergent when $z\rightarrow-\infty$. The solutions for $G_n(x)$ are
determined by
\begin{eqnarray}
\label{AppD-G}
G_{n}(x)=\frac{\omega}{\lambda_n}\left[\frac{m}{\omega}s\frac{x-\frac{a}{b}}{x-1}F_{n}(x)-x\frac{d}{dx}F_{n}(x)\right].
\end{eqnarray}
The solutions (\ref{AppD-1})-(\ref{AppD-G}) are well behaved when
$\mu_n>0,~\rho \geq 1$. They are satisfied in our numerical example
in Sec.~\ref{sec:3}.

We find that Eq.~(\ref{suppose2-4}) also has a well behaved zero
mode solution. This zero mode solution is given by
\begin{eqnarray}
\label{AppD-Z}
F_0(x)&=&\frac{\sqrt{\omega}}{\sqrt{N_0}}x^{\frac{m}{\omega}s}(1-x)^{\rho},~G_0(x)=0.
\end{eqnarray}
It is consistent with what we get from (\ref{ZeroMode}).

\section{Another metric example for orthogonality}\label{sec:E}

In this appendix, we suggest another metric which ensures that the
the orthogonal conditions (\ref{norm2a}) are satisfied. This metric
is given by
\begin{eqnarray}
\label{suppose3} B(z)=s \frac{\omega z-a}{\omega z-b},~~s,a,b,\omega
>0.
\end{eqnarray}
We suppose $a,b>0$ and $a>b$ here. The conditions $a,b<0$ and $a<b$
work well also. But the conditions $a<0,b>0$ do not work.
Eqs.~(\ref{solution15}) and (\ref{solution16}) can be solved by
confluent hypergeometrical function~(or Kummer's function). When
$\gamma=2(1+\nu),\nu=\frac{m}{\omega}s(a-b)$ is not integer, the
solutions are given by
\begin{eqnarray}
\label{suppose3-1} F(z)&=&C_1 e^{-\mu (\omega z-b)}(\omega
z-b)^{1+\nu} \mathrm{hypergeom}\left(\alpha;\gamma,2\mu(\omega
z-b)\right)\nonumber \\
&+&C_2 e^{-\mu (\omega z-b)}(\omega z-b)^{1+\nu+1-\gamma}
\mathrm{hypergeom}\left(\alpha+1-\gamma;2-\gamma,2\mu(\omega
z-b)\right),
\end{eqnarray}
in which
$\mu=\sqrt{\left(\frac{m}{\omega}s\right)^2-\left(\frac{\lambda}{\omega}\right)^2}$
and $\alpha=1+\nu-\frac{m}{\omega}s \frac{\nu}{\mu}$. We choose the
range of $z$ to be $[R,\infty)$, where $R=\frac{b}{\omega}$. The
confluent hypergeometrical function $F(\alpha;\gamma,\xi)\sim
e^{\xi}$ when $\xi\rightarrow \infty$. In order to make the
solutions well behaved when $z\rightarrow\infty$, the confluent
hypergeometrical function must be cut off to be a polynomial by the
requirement
\begin{eqnarray}
\label{suppose3-2} \alpha= 1+\nu-\frac{m}{\omega}s
\frac{\nu}{\mu}=-(n-1),~n=1,2,\cdots.
\end{eqnarray}
Then $\lambda_n$ are determined to be
\begin{eqnarray}
\label{suppose3-3}
\lambda_n=\left[\frac{m}{\omega}s\left(1-\frac{\nu^2}{(\nu+n)^2}\right)^{\frac{1}{2}}\right]\omega.
\end{eqnarray}
By (\ref{suppose3-3}), we know that
\begin{eqnarray}
\label{suppose3-4}
\left[\frac{m}{\omega}s\frac{(2\nu+1)^{\frac{1}{2}}}{\nu+1}\right]\omega\leq\lambda\leq
\left(\frac{m}{\omega}s\right) \omega.
\end{eqnarray}

We should also require that the solutions are well behaved at
$z=R=\frac{b}{\omega}$. This condition requires the solutions
further to be
\begin{eqnarray}
\label{suppose3-5} F_n(z)&=&\frac{\sqrt{\omega}}{\sqrt{N_1}} e^{-\mu
(x-b)}(x-b)^{1+\nu}
\mathrm{hypergeom}\left(-n;\gamma,2\mu(x-b)\right).
\end{eqnarray}
Here we define $x=\omega z$, and $G_n(z)$ are determined by
\begin{eqnarray}
\label{AppE-G}
G_{n}(x)=\frac{\omega}{\lambda_n}\left[\frac{m}{\omega}s\frac{x-a}{x-b}F_{n}(x)-\frac{d}{dx}F_{n}(x)\right].
\end{eqnarray}
There is also a well behaved zero mode solution, which is given by
\begin{eqnarray}
\label{AppE-1}
F_0(x)=0,~G_0(x)=\frac{\sqrt{\omega}}{\sqrt{N_0}}e^{-s\frac{m}{\omega}
x}(x-b)^{\nu}.
\end{eqnarray}
We have supposed $a>b$. So when
$z\rightarrow\infty,x\rightarrow\infty$, $F_n(z),G_n(z)\rightarrow
0$; while $F_n(z),G_n(z)=0$ at $z=R, x=b$ if $\nu\geq 2$. So the
orthogonal conditions (\ref{norm2b-1}) can be satisfied.

According to the analysis in \cite{Hosotani:2006-2}, if we let
$\lambda_0=m s$ to be the lightest generation of SM, there exist
infinite heavier generations corresponding to $\lambda < m s$, in
which $\lambda=0$ is the heaviest generation. However, there exists
a problem in this case: the lighter generations approximate to be
continuous, which conflicts with the experimental fact. Therefore,
special boundary conditions must be adopted to remove the reductant
generations. We hope that there are only finite generations left
with the help of the special boundary conditions. But it seems that
it is difficult to impose such boundaries naturally.

\section{Ricci scalar curvature for metric}\label{sec:F}

Ricci scalar curvature for metric Ansatz (\ref{ansatz1}) is given by
\begin{eqnarray}
\label{AppF-1}
R=-\left[10(B^{-3}B_{zz}+B^{-4}B_{z}^2)+(8A^{-3}A_{yy}+4A^{-4}A_{y}^2)B^{-2}\right],
\end{eqnarray}
in which $A_y=\frac{dA}{dy},~B_z=\frac{dB}{dz}$. From the second
term in (\ref{AppF-1}), we know that there is singularity for the
metric (\ref{suppose2}) at $z=\frac{\mathrm{log}a}{\omega}$ and the
metric (\ref{suppose3}) at $z=\frac{a}{\omega}$. At the point
$z=\frac{\mathrm{log}b}{\omega}$ and $z=\frac{b}{\omega}$, the
metric (\ref{suppose2}) and (\ref{suppose3}) are not well defined
respectively, but the Ricci scalar for them are well defined. So it
needs further work to determine whether they are true singularities.


\end{document}